# Local characterization and engineering of proximitized correlated states in graphene-NbSe$_2$ vertical heterostructures

Zhiming Zhang[1], Kenji Watanabe[2], Takashi Taniguchi[3], Brian J. LeRoy*[1]

[1]*Physics Department, University of Arizona, 1118 E 4th Street, Tucson, AZ 85721, USA.*
[2]*Research Center for Functional Materials, National Institute for Materials Science, Namiki 1-1, Tsukuba, Ibaraki 305-0044, Japan.*
[3]*International Center for Materials Nanoarchitectonics, National Institute for Materials Science, Namiki 1-1, Tsukuba, Ibaraki 305-0044, Japan.*

(Dated 2020)

Using a van der Waals (vdW) vertical heterostructure consisting of monolayer graphene, monolayer hBN and NbSe$_2$, we have performed local characterization of induced correlated states in different configurations. At a temperature of 4.6 K, we have shown that both superconductivity and charge density waves can be induced in graphene from NbSe$_2$ by proximity effects. By applying a vertical magnetic field, we imaged the Abrikosov vortex lattice and extracted the coherence length for the proximitized superconducting graphene. We further show that the induced correlated states can be completely blocked by adding a monolayer hBN between the graphene and the NbSe$_2$, which demonstrates the importance of the tunnel barrier and surface conditions between the normal metal and superconductor for the proximity effect.

## I. INTRODUCTION

Because of the unique geometry and electronic structure of graphene [1], recently there has been a significant interest on inducing correlated states such as superconductivity in this relativistic quantum system [2–11]. When a normal metal is placed in good contact with a superconductor, Cooper pairs can be induced in the normal metal through the proximity effect [12–14]. Interesting physics such as specular Andreev reflection [15–18], Klein-like tunneling [19] and the interplay between Andreev states with quantum hall states [20,21] has been observed in proximitized superconducting graphene systems.

One of the ideal candidates for making a graphene-superconductor junction is NbSe$_2$, a two dimensional material with both superconductivity and charge density wave (CDW) transitions at low temperatures [22,23]. Although several electrical transport experiments have already been performed with graphene-NbSe2 heterostructure devices [17,21,24–26], there is still a lack of local spectroscopic information for this heterostructure. In this study, we use scanning tunneling microscopy and spectroscopy to directly probe the superconducting gap, doping level, CDWs and vortex lattices in a graphene-NbSe$_2$ vertical heterostructure. Furthermore, with the insertion of a monolayer hBN (MLhBN) between the hBN and NbSe$_2$, we have found that the correlated states can be completely blocked.

## II. EXPERIMENTAL DETAILS

To fabricate our device, graphene and hBN were mechanically exfoliated from bulk crystals and deposited on 285 nm and 90 nm thick SiO$_2$ wafers respectively. The MLhBN was identified under an optical microscope with 590 nm monochromatic light to optimize the contrast [27]. The NbSe$_2$ flake with a thickness of ~45nm was exfoliated inside a glovebox environment with oxygen level < 1ppm. The vdW heterostructure was created with a dry transfer technique [28] inside the glovebox and the NbSe$_2$ is encapsulated by the graphene and a thick hBN flake to prevent it from oxidizing. The heterostructure was fabricated such that the MLhBN partially covered the NbSe$_2$ giving a region where graphene was in direct contact with NbSe$_2$ and another region where they were separated by a monolayer of hBN. 5nm-Cr/ 50nm-Au contacts were created with electron-beam lithography and physical vapor deposition. The optical image of the completed device is shown in Fig. 1(a), where the gray and blue dashed lines indicate the MLG and MLhBN respectively.

STM/STS measurements were performed in an ultrahigh-vacuum LT-STM (Omicron) operating at 4.6 K, Fig. 1(b) shows a schematic of the experimental setup. dI/dV spectroscopies were acquired by adding 0.4~5 mV modulation voltages (V$_{mod}$) at a frequency of 617 Hz to the bias voltage and measuring the current with lock-in detection. All the tips were first checked on the Au surface to ensure that they had the proper work function based on the decay of the tunnel current with distance from the sample. In addition, dI/dV spectroscopy was performed on the Au surface to ensure that the tip had a constant density of states. A small perpendicular magnetic field was applied to the device by mounting the sample on top of a permanent magnet (D43-N52, K&J Magnetics).

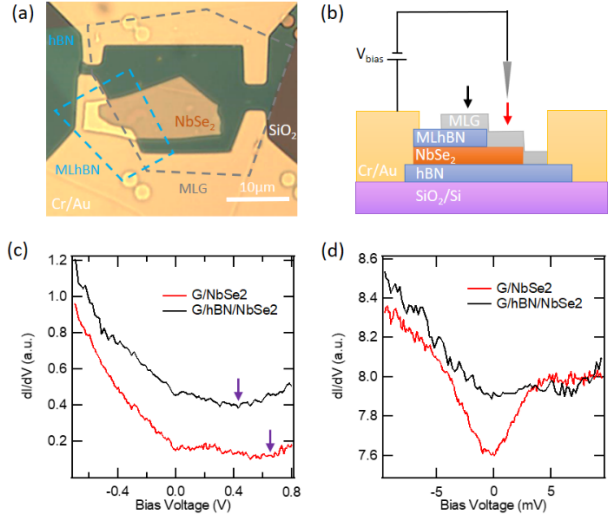

FIG. 1. (a) Optical microscopy image of the measured device. Gray and blue dashed lines enclose the monolayer graphene and monolayer hBN flakes. (b) Schematic of the STM experimental setup, black and red arrows indicated the position where the dI/dV curves in (c) and (d) were taken. (c) dI/dV spectra acquired with I = 100pA, $V_{mod}$ = 5mV. (d) dI/dV spectra acquired with I = 500pA, $V_{mod}$ = 0.4mV.

## III. RESULTS AN DISCUSSION

### A. Dirac point and superconducting gap

Fig. 1(c) shows dI/dV spectra on the two different stacking configurations as indicated by the black and red arrows in Fig. 1(b). For both areas, the spectra show an overall V-shaped graphene density of states feature and the graphene is hole-doped. The Dirac point of the graphene is at ~0.65 V in the graphene/NbSe$_2$ area, and ~0.43 V in the graphene/hBN/NbSe$_2$ area, as indicated by the purple arrows. This is because the MLhBN lowers the work function [29] of the heterostructure under the graphene, making the graphene less p-doped. There is also a flattening of the density of states from the Fermi level to ~0.2 V for the red curve but not for the black curve, this indicate that the band structure of the graphene is being modified more by the NbSe$_2$ in the graphene/NbSe$_2$ area.

Fig. 1(d) shows high resolution spectroscopy on the two stacking configurations near the Fermi level. There is a soft gap opened near the Fermi level in the graphene/NbSe$_2$ area but not in the graphene/hBN/NbSe$_2$ area, indicating that the graphene directly sitting on the NbSe$_2$ area becomes superconducting as predicted by theory [30], while the graphene remains normal when there is the MLhBN between the graphene and the superconducting NbSe$_2$. From the tunneling model of the superconducting proximity effect [13], the induced superconductivity depends on the barrier height between the superconductor and the normal metal. In our case, the insertion of a MLhBN not only induces an additional atomic layer of hBN but also creates different interfaces between the materials, thus greatly increasing the barrier height between the graphene and NbSe$_2$ and making the induced gap not observable under our experimental conditions.

### B. Determination of stacking configurations

By taking high resolution topography images of different areas of the device, we can determine the stacking orientations from the moiré pattern formed between the different lattices. Figures 2(a), (c), and (e) show topography images of the three different stacking configurations, graphene on NbSe$_2$ (G/NbSe$_2$), graphene on MLhBN on NbSe$_2$ (G/hBN/NbSe$_2$) and MLhBN on NbSe$_2$ (hBN/NbSe$_2$). Figures 2(b), (d), and (f) are the Fourier transforms of the corresponding topography images. Due to the hexagonal symmetry of the lattices, we have employed a six-fold symmetrization procedure [31] to increase the signal-to-noise ratio in our Fourier transforms. Blue hexagons and orange rectangles mark the graphene and NbSe$_2$ lattices respectively. Colored triangles mark the moiré superlattices formed by the three different possible combinations of two lattices. The wavelength of the moiré pattern is given by:

$$\lambda = \frac{(1+\delta)a}{\sqrt{2(1+\delta)(1-\cos\phi) + \delta^2}} \quad (1)$$

where $a$ is the shorter lattice constant of the two lattices, $\delta$ and $\phi$ are the lattice mismatch and the twist angle between the two lattices. The relative angle θ of the

moiré pattern with respect to the shorter lattice is given by:

$$\tan\theta = \frac{\sin\phi}{(1+\delta) - \cos\phi} \qquad (2)$$

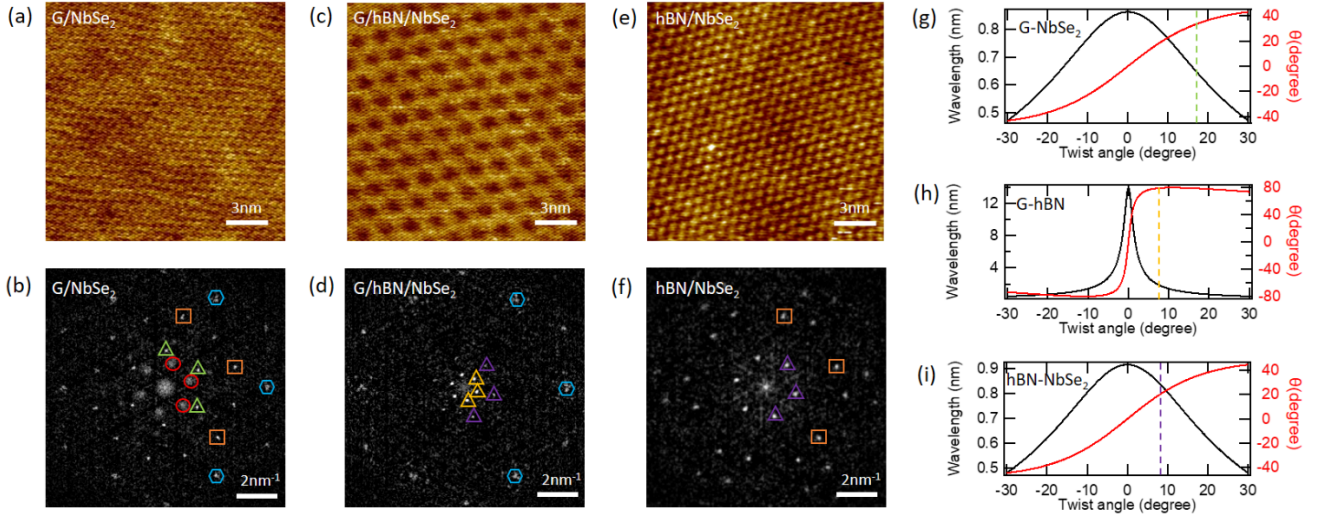

FIG. 2. (a), (c), (e): Topography images of the three different stacking configurations, acquired with $V_{bias}$ = 0.3 V, I = 100 pA. (b), (d), (f): Symmetrized Fourier transform of (a), (c) and (e). Blue hexagons and orange rectangles mark the graphene and $NbSe_2$ lattices; red circles mark the charge density waves; green, yellow and purple triangles mark the graphene-$NbSe_2$ moiré, graphene-hBN moiré and hBN-$NbSe_2$ moiré respectively. (g)-(i): Theory calculation of moiré wavelengths and $\theta$ for three different configurations, dashed lines indicate the experimental values.

Figures 2(g)-(i) plot the wavelength and $\theta$ as a function of twist angle for all three possible combinations of two lattices. By measuring $\lambda$ and $\theta$ of the moiré pattern, we can determine which two of the lattices formed the moiré as well as the twist angle between the lattices, the corresponding data points obtained from the Fourier transform are labeled as colored dashed lines. The fact that graphene-$NbSe_2$ moiré pattern only shows up in the graphene/$NbSe_2$ area but not in graphene/hBN/$NbSe_2$ area indicates that the MLhBN blocks the strong electronic coupling between the graphene and the $NbSe_2$.

Red circles in Fig. 2(b) mark the charge density waves (CDWs), which have similar feature as the CDWs that have been observed in $NbSe_2$ [32]: disks in the Fourier transform are centered at three times the wavelength of the $NbSe_2$ lattice. Such features are not obvious in the hBN/$NbSe_2$ area and not observable in the graphene/hBN/$NbSe_2$ area, indicating that the CDWs can be induced in graphene when the graphene is sitting directly on the $NbSe_2$, while the characteristics of the CDWs are not preserved when the electrons are tunneling through MLhBN.

### C. Vortices in graphene on $NbSe_2$

To further study the properties of the induced superconductivity in graphene, we apply a 0.26 T magnetic field perpendicular to the sample and investigate the vortices that form in the G/$NbSe_2$ area. Figure 3(a) shows a local density of states (LDOS) map measured by fixing $V_{bias}$ at -3 mV and scanning over the sample area while recording dI/dV as a function of real space position. From the image, we can clearly see the emergence of Abrikosov vortices [33], providing further evidence that the superconductivity is induced in the graphene by the underlying $NbSe_2$.

Fig. 3(b) shows multiple dI/dV spectroscopies measured at different distances from the center of the vortex along the line indicated by the blue arrow in Fig. 3(a), the superconducting gap centered around the Fermi level becomes smaller and the quasiparticle peak at around 4 mV is weaker when it is closer to the center of the vortex. The asymmetry of the spectra is from the fact that the Dirac point is higher in energy than the Fermi level.

To see how the superconducting gap changes as a function of distance from the vortex center, we fit each dI/dV spectroscopy curve and extract the superconducting gap. Because of the asymmetric nature of the measured dI/dV curves, only the data

above the Fermi level were used for the fitting. At zero temperature, the Dynes formula [34] is given by:

$$\rho(E, \Gamma) = \rho_0 \text{Re}\left[\frac{E - i\Gamma}{\sqrt{(E - i\Gamma)^2 - \Delta^2}}\right] \quad (3)$$

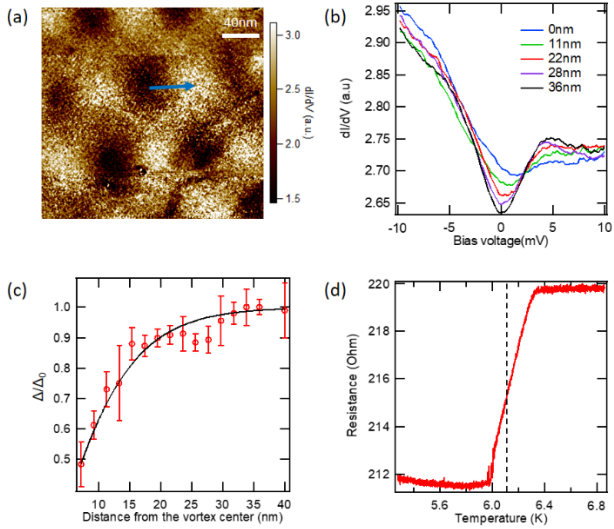

FIG 3. (a) dI/dV map showing the vortices in graphene/NbSe$_2$ area, acquired with V$_{bias}$ = -3 mV, I = 200 pA, V$_{mod}$ = 0.4 mV. Blue arrow indicates the position where the line cut spectroscopy were taken. (b) dI/dV spectra at different distances from the center of a vortex, acquired with I = 500 pA, V$_{mod}$ = 0.4 mV. (c) Extracted superconducting gap plotted against the distance from the vortex center, black curve indicates the fitting function. (d) Two terminal resistance measurement as a function of temperature, dashed line corresponding to the critical temperature.

where $\rho$ is the density of states, $\rho_0$ is the normal-state density of states at the Fermi level, $\Gamma$ accounts for the broadening effects other than temperature. To include the finite temperature effects, we integrate the density of states with the derivative of the Fermi-Dirac distribution $f$, the measured density of states $N$ is then given by:

$$N(V) = N_0 \int_{-\infty}^{\infty} dE \left(-\frac{\partial f}{\partial E}\right) \rho(E + eV, \Gamma) \quad (4)$$

Fig. 3(c) shows the extracted superconducting gap as a function of the distance from the vortex center $r$. The superconducting gap far away from the vortex $\Delta_0$ was determined by the two-terminal temperature dependent resistance measurement shown in Fig. 3(d). We define the measured critical temperature $T_C$ as the midpoint of the step transition, then $\Delta_0$ was calculated by using the equation [35]: $\Delta_0 = 1.764 k_B T_C$. For our device, we have obtained that $T_C \sim 6.1 K$ and $\Delta_0 = 0.93$ meV, which is ~85% percent of the value for a bulk NbSe$_2$ crystal [22]. This ratio describes the quality of the interface between the normal graphene and the superconducting NbSe$_2$. Our reduction in T$_c$ is comparable to another experiment when aluminum was directly deposited on the graphene [18], indicating that a high quality interface was achieved by our sample fabrication process.

Another parameter that can represent the barrier strength between a type II superconductor and normal metal is the coherence length $\xi$, which is expected to increase for decreasing interface transparency [36]. We use the following equation [37] to obtain the coherence length from the extracted superconducting gap:

$$\frac{\Delta}{\Delta_0}(r) = \tanh\left(\frac{r}{\xi}\right) \quad (5)$$

from the fitting curve in Fig. 3(d) we have obtained that $\xi = (13.5 \pm 0.5)$ nm, which is greater than the coherence length in bulk NbSe$_2$ [38], suggesting a lower upper critical filed in graphene-NbSe$_2$ heterostructure than in bulk NbSe$_2$.

### D. Scattering waves

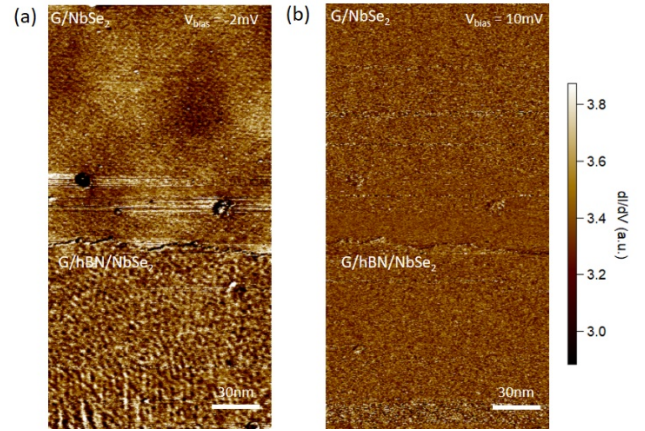

FIG 4. (a) dI/dV map under a perpendicular magnetic field around the monolayer hBN edge, acquired with V$_{bias}$ = -2 mV, I = 200 pA, V$_{mod}$ = 0.4 mV. (b) Same image as (a) except acquired with V$_{bias}$ = 10 mV.

In Fig. 1(d) we have shown that the superconducting gap is not present for the G/hBN/NbSe$_2$ area, we further confirm this by imaging the LDOS near the MLhBN edge in the presence of an external magnetic field, as shown in Fig. 4, the upper area is the G/NbSe$_2$ area and the lower area is the G/hBN/NbSe$_2$ area. When imaging close to the superconducting gap, V$_{bias}$ = -2 mV, from Fig. 4(a) we can see that the Abrikosov vortices are only present in the upper graphene/NbSe$_2$ area, consistent with our spectroscopic data. Additionally, we observed long-wavelength scattering waves in the lower graphene/hBN/NbSe$_2$ area, similar to those observed in graphene near an atomic step edge [39,40] or near defects [41]. When imaging at higher voltage, V$_{Bias}$ = 10 mV, both the vortices and the scattering waves are gone, as shown in Fig. 4(c), since the amplitude of the scattering waves in graphene decay very fast with energy [39].

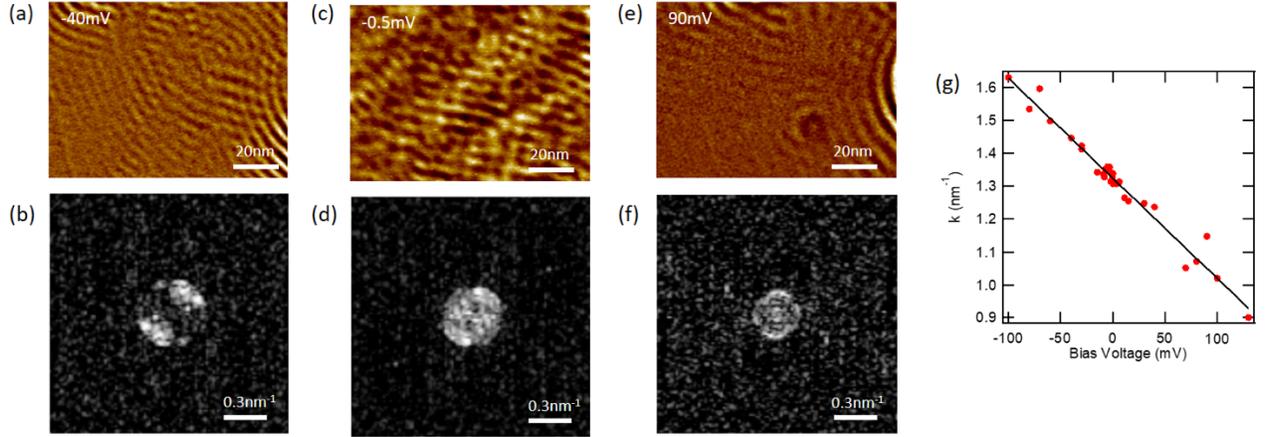

FIG. 5. (a) dI/dV map near surface defects, acquired with V$_{bias}$ = -40 mV, I = 500 pA, V$_{mod}$ = 3 mV. (b) dI/dV map acquired with V$_{bias}$ = -0.5 mV, I = 50 pA, V$_{mod}$ = 0.4 mV. (c) dI/dV map acquired with V$_{bias}$ = 90 mV, I = 40 pA, V$_{mod}$ = 5 mV. (b), (d), (f): Fourier transform of (a), (c), (e). (g) Wave vector of the scattering wave as a function of bias voltage, solid black line indicates the fitting function.

The scattering waves can be used to determine the dispersion of the material. We measure the LDOS maps at different energies in an area close to many surface defects so that the scattering waves are strong. Figures 5(a), (c), and (e) are selected LDOS images taken at negative tip voltage, close to the Fermi level and positive tip voltage. Figures 5(b), (d), (f) are the Fourier transforms of the above images, the disk-like feature at the center is due to intravalley scattering process [31]. Its size shrinks as the wavelength of electrons becomes longer and therefore by measuring its diameter as a function of tip voltage, we can obtain the energy versus momentum dispersion relation. Figure 5(g) shows the wavevectors of the scattering waves measured from the Fourier transform images, as expected from the graphene band structure, it can be fit with a linear equation [39]:

$$k = \frac{2}{\hbar v_f}(eV_{bias} - eV_0) \quad (5)$$

where $e$ is the charge of an electron, $V_0$ is the position of the Dirac point, $v_f$ is the fermi velocity of the electrons. From the fitting we obtained that $V_0 = (437 \pm 15)$ meV, consistent with our spectroscopy data in Fig. 1(c), $v_f = (1.00 \pm 0.03) \times 10^6$ m/s, consistent with theory [1].

## IV. CONCLUSIONS

In summary, we have found that both the CDWs and proximitized superconductivity exist in the graphene-NbSe$_2$ heterostructure. By applying a magnetic field, we directly imaged the Abrikosov vortices in the graphene/NbSe$_2$ area and extracted the coherence length from the fitted superconducting gaps. Furthermore, by inserting a MLhBN between the graphene and the NbSe$_2$, both the CDWs and superconductivity are suppressed in graphene, which demonstrates the importance of the barrier strength between the normal metal and superconductor interface for proximitized effects. From the scattering waves, we have obtained the dispersion relation of the graphene on the G/MLhBN/NbSe$_2$ substrate, which is consistent with our spectroscopic study and the theory [1]. The above observations indicate that even a single monolayer of hBN is a very good barrier to block interactions between the graphene and the NbSe$_2$.

Our experiment is the first local characterization of the graphene-NbSe$_2$ heterostructure. We have demonstrated the importance of the barrier strength for the proximitized correlated states including CDWs and superconductivity in vdW heterostructures. Moreover, we provide an innovative way to engineer the proximitized correlated states by the insertion of MLhBN, which opens the possibility of making more versatile superconducting devices and circuits in the future.

## ACKNOWLEDGEMENTS

The work at the University of Arizona was supported by the National Science Foundation under grant DMR-1708406 and the Army Research Office under Grant No. W911NF-18-1-0420. K.W. and T.T. acknowledge support from the Element Strategy Initiative conducted by the MEXT, Japan, Grant Number JPMXP0112101001, JSPS KAKENHI Grant Numbers JP20H00354 and the CREST(JPMJCR15F3), JST.

---